\documentclass[conference,a4paper]{APSIPA2021}

\usepackage[hyphens]{url}

\usepackage{
amsmath,
graphicx,
booktabs,
footmisc,
}

\usepackage[ruled,vlined]{algorithm2e}

\usepackage{hyperref}
\hypersetup{
colorlinks=true,
linkcolor=blue,
citecolor=blue,
urlcolor=blue,
}

\begin{document}

\title{Generation of Speaker Representations Using Heterogeneous Training Batch Assembly}

\author{%
\authorblockN{%
Yu-Huai Peng, Hung-Shin Lee, Pin-Tuan Huang, and Hsin-Min Wang
}
\authorblockA{%
Institute of Information Science, Academia Sinica, Taiwan}
}

\maketitle
\thispagestyle{empty}

\begin{abstract}
In traditional speaker diarization systems, a well-trained speaker model is a key component to extract representations from consecutive and partially overlapping segments in a long speech session. To be more consistent with the back-end segmentation and clustering, we propose a new CNN-based speaker modeling scheme, which takes into account the heterogeneity of the speakers in each training segment and batch. We randomly and synthetically augment the training data into a set of segments, each of which contains more than one speaker and some overlapping parts. A soft label is imposed on each segment based on its speaker occupation ratio, and the standard cross entropy loss is implemented in model training. In this way, the speaker model should have the ability to generate a geometrically meaningful embedding for each multi-speaker segment. Experimental results show that our system is superior to the baseline system using x-vectors in two speaker diarization tasks. In the CALLHOME task trained on the NIST SRE and Switchboard datasets, our system achieves a relative reduction of 12.93\% in DER. In Track 2 of CHiME-6, our system provides 13.24\%, 12.60\%, and 5.65\% relative reductions in DER, JER, and WER, respectively.
\end{abstract}

\section{Introduction}
Speaker diarization is known as a ``who spoke when'' problem. It aims to group together speech segments produced by the same speaker within an audio stream \cite{Tranter2006}. It has been studied in various application scenarios, such as telephone conversations, broadcast news, meeting recordings, and daily life conversations \cite{Barras2006,Sell2018}. For automatic speech recognition (ASR) in real-life multi-speaker scenarios, it is an indispensable pre-processing for the recorded audio. Without this pretreatment, ASR performance may be severely degraded. For example, we can see this in the results of Track 2 of the CHiME-6 Challenge\footnote{\url{https://chimechallenge.github.io/chime2020-workshop/programme.html}}. In recent years, several end-to-end methods have been proposed and realized by probabilistic models \cite{Diez2018} and neural networks \cite{Fujita2019a,Fujita2019b}. However, in this paper, we will focus on improving the state-of-the-art two-stage approach by enhancing speaker modeling.

After removing the non-speech part of an audio stream, the operation of a typical speaker diarization system can be divided into two stages: segmentation and clustering \cite{Cheng2009}. In the segmentation stage, the entire speech session is uniformly cut into a sequence of half-overlapping short segments. Each segment is transformed into a vector representation that contains the discriminative information of the speaker, such as i-vector \cite{Dehak2011,Sell2014}, x-vector \cite{Sell2018,Snyder2018,Diez2019}, and d-vector \cite{Wan2018,Wang2018,Wang2019}. The models used to generate these segment embeddings are usually trained from a large amount of acoustic frames labeled with speaker identities. In the clustering stage, based on the similarity calculation provided by probabilistic linear discriminant analysis (PLDA) or cosine measure \cite{Sell2014,Ioffe2006}, segments with homogeneous representations are grouped into the same cluster by a suitable algorithm. Many clustering approaches, such as agglomerative hierarchical clustering (AHC) \cite{Sell2014,Garcia-Romero2017}, spectral clustering \cite{Wang2018,Ning2006,Park2019b}, Gaussian mixture
model (GMM) \cite{Shum2013}, hidden Markov model (HMM) \cite{Diez2019}, and unbounded interleaved-state recurrent neural networks (UIS-RNN) \cite{Wang2019}, have been applied to speaker diarization. In addition, VB-diarization (VBD), an advanced algorithm based on Variational Bayes, has proven to be effective in further fine tuning the speaker diarization results \cite{Sell2015}.

However, there is still room for improvement in the two-stage framework. In most approaches, the speaker representation extractor, also called the speaker model, is borrowed from the speaker verification task. The speaker model is usually trained on corpora, such as the NIST SRE-2000\footnote{\url{https://catalog.ldc.upenn.edu/LDC2001S97}\label{LDC2001S97}} and VoxCeleb-2 datasets \cite{SonChung2018}, which have long speech segments, and each segment contains a single speaker. In speaker diarization, nevertheless, the short segments given by uniform segmentation, which are less than 2 seconds in general, may contain more than one speaker so that the resulting speaker embeddings may contain characteristics of out-of-segment speakers. In addition, 2 seconds may be too short to infer a reliable speaker embedding for clustering. In summary, there is a mismatch between training and inference phases.

Post-processing methods such as traditional re-segmentation \cite{Meignier2006,Kenny2010} and advanced VBD \cite{Diez2018,Sell2015} can remedy the above problem to some extent, thereby improving the speaker diarization performance. Alternatively, in order to make the front-end speaker model more objective-driven for the speaker diarization task, we can divide a long training session into shorter segments of appropriate length and consider the heterogeneity of speakers in a single segment in speaker modeling. Therefore, in this paper, we propose a novel scheme for developing the speaker model to reduce the mismatch when extracting the speaker representations. Our scheme aims to make the speaker model better handle segments containing multiple speakers and make the speaker representations more reliable. For example, our model guarantees to a certain extent that the vector representation of a segment that mixes two speakers is closer to either of the two speakers than any other speaker. The experimental results in two speaker diarization tasks show that our system is superior to the baseline using x-vectors. In the CALLHOME task trained on the NIST SRE and Switchboard datasets, our system achieves a relative reduction of 12.9 \% in diarization error (DER). In Track 2 of CHiME-6, our system achieves relative reductions of 13.24\%, 12.60\%, and 6.57\% in DER, Jaccard error rate (JER), and word error rate (WER), respectively.

This paper is organized as follows. Section \ref{system_overview} presents the procedures of two baseline systems for speaker diarization. Section \ref{proposed_model} explains the proposed speaker modeling method with some illustrations. The experiment settings and results are described and discussed in Section \ref{experiments}. Section \ref{conclusions} gives the conclusions and future work.

\section{System Overview}
\label{system_overview}
In this paper, we implemented two baseline speaker diarization systems, where the speaker models were based on the time delayed neural networks (TDNNs, also known as the x-vectors) according to Kaldi's recipe\footnote{\url{https://github.com/kaldi-asr/kaldi/tree/master/egs/callhome_diarization/v2}} and the ResNet-34 realized by PyTorch \cite{Cai2018a}, respectively. Our proposed speaker model training scheme was developed on top of the second baseline.

\subsection{Speaker representation extraction}
The x-vector is a kind of speaker representation produced by TDNNs \cite{Snyder2018}. It was originally proposed for the speaker verification task, and was later applied directly to the speaker diarization task without any modification, even without retraining \cite{Sell2018,Diez2019}. In both tasks, the x-vector is used to describe the speaker characteristics of a speech segment. By calculating the similarity between two x-vectors, we can estimate the degree to which two speech segments are produced by the same speaker.

For the convenience of implementation, we used the convolutional neural network (CNN) based architecture similar to that described in \cite{Cai2018a} for speaker modeling in our proposed scheme. As shown in Table \ref{tab:resnet}, after feed-forwarding through 5 ResNet-based layers (\texttt{res1} to \texttt{res5}) and an operation of average pooling, a 256-dimensional speaker embedding was derived in the \texttt{linear1} layer while the \texttt{linear2} layer was the penultimate layer for speaker classification. In this way, we implemented the second baseline speaker diarization system by replacing the x-vector structure with the ResNet-34 speaker model in order to apply our new training scheme later. This speaker model is hereafter referred to as ``ResNet-34''.

In the training stage, the speaker models were trained with a large amount of utterances labeled with speaker identities. To train the x-vector model, the utterances with speaker labels were chopped into segments of 2 to 4 seconds, then the Mel-frequency cepstral coefficients (MFCCs) were extracted with the frame length of 0.025 seconds and the frame shift of 0.01 seconds as the input. The dimensionality of MFCCs was set to 23 in the CALLHOME task and 30 in the CHiME-6 task. To train the ResNet-34 model, the utterances were chopped into segments of 3 to 8 seconds, then the 64-dimensional Mel-filter banks were extracted with the frame length of 0.025 seconds and the frame shift of 0.01 seconds as the input. All the features were subjected to mean subtraction with a sliding window of 300 frames before being input to the speaker model.

\subsection{Clustering}
After segmentation and speaker representation extraction, the PLDA model was used to score the similarity between two speaker representations. The PLDA model was trained with the same corpus used to train the speaker models. The speaker representation was first extracted from each speech segment of 3 seconds, and then subjected to centering, whitening and length normalization. Finally, the speaker loading matrix was learned via the expectation maximization (EM) algorithm.

In the inference stage, the speaker representation of each speech segment was extracted first. Then, the similarity between each pair of segments was calculated by the PLDA model to get the similarity matrix. In this way, each row vector of the similarity matrix was used to represent the corresponding speech segment. Finally, the AHC algorithm was applied to the clustering of row vectors. If the number of speakers in a speech session is known, the merging process ends when the number of clusters is equal to the number of speakers. Otherwise, the stopping criterion for merging is determined based on the development set.

\begin{table}
\caption{The architecture and specifications of ResNet-34, where \texttt{res} denotes the ResNet-based layer, and $B$, $T$, $D$, and $C$ represent the batch size, temporal length, feature dimensionality, and number of training speakers, respectively.}
\vspace{-5pt}
\label{tab:resnet}
\renewcommand\arraystretch{1.5}
\centering
\begin{tabular}{ccccc}
\toprule
Layer & Feature Size & Downsample & \# Blocks \\ 
\midrule\midrule
\texttt{input} & $B \times 1 \times T \times D$ & - & - \\
\midrule
\texttt{conv} & $B \times 16 \times T \times D$ & False & - \\
\midrule
\texttt{res1} & $B \times 16 \times T \times D$ & False & 3 \\
\midrule
\texttt{res2} & $B \times 32 \times \frac{T}{2} \times \frac{D}{2}$ & True & 4 \\
\midrule
\texttt{res3} & $B \times 64 \times \frac{T}{4} \times \frac{D}{4}$ & True & 6 \\
\midrule
\texttt{res4} & $B \times 128 \times \frac{T}{8} \times \frac{D}{8}$ & True & 3 \\
\midrule
\texttt{res5} & $B \times 256 \times \frac{T}{16} \times \frac{D}{16}$ & True & 3 \\
\midrule
\texttt{pooling} & $B \times 256$ & - & - \\
\midrule
\texttt{linear1} & $B \times 256$ & - & - \\
\midrule
\texttt{linear2} & $B \times C$ & - & - \\
\bottomrule
\end{tabular}
\vspace{-5pt}
\end{table}

\subsection{Re-segmentation}
The speech session input to the speaker diarization system was first chopped into a sequence of short segments using a sliding window of 1.5 seconds (50\% overlap). Then, the speaker representation (e.g., the x-vector) of each segment was extracted by the speaker model. Next, the PLDA model and the AHC algorithm were used for clustering. After segmentation and clustering, the preliminary diarization result was obtained.

The VBD algorithm was applied to fine-tune the preliminary diarization result. In the training phase, a 1024-component UBM-GMM with diagonal covariance matrices and 400 eigenvoice bases were trained with 30-dimensional MFCCs. In the inference stage, the preliminary diarization result was used as the initialization in the VBD model. The tunable parameters were determined based on the development set.

\begin{figure}[!t]
\centering
\includegraphics[width=0.45\textwidth]{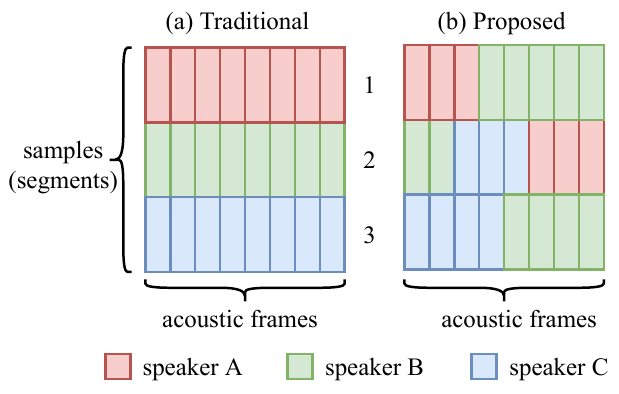}
\vspace{-5pt}
\caption{Illustration of the traditional and our proposed training schemes with respect to the speaker distribution in a mini-batch.}
\label{fig:batch}
\vspace{-10pt}
\end{figure}

\section{Proposed Speaker Modeling}
\label{proposed_model}
As mentioned above, there are some limitations with conventional speaker models. When these speaker models are applied to speaker diarization tasks, they may perform poorly on segments containing more than one speaker. Therefore, inspired by \cite{Fujita2020}, we propose to improve ``speaker modeling'' by augmenting the training data with segments containing multiple speakers without any overlapping regions. In \cite{Fujita2020} (i.e., EEND for short), the authors simulated overlapping speech by randomly mixing two utterances of different speakers ``for'' making the mixed utterance correspond to two label sequences of speaker-dependent voice activity detection (VAD), i.e., the diarization results. However, in EEND, the concept of speaker embeddings is not apparent, nor the homogeneity or heterogeneity of speaker embeddings can be clearly described. 

In our method, a soft label is applied to each segment (sample) based on the speaker occupation ratio, and the standard cross entropy loss is used in model training. Take Figure \ref{fig:batch} as an example. As shown in Figure \ref{fig:batch} (a), traditionally, the speaker model is trained with a large amount of speech segments each containing only one speaker. Therefore, the ground truths of the training samples 1, 2, and 3 are $[1,0,0]$, $[0,1,0]$, and $[0,0,1]$, respectively, if there are three training speakers A, B, and C. In contrast, we artificially compose a large number of speech fragments each containing more than one speaker, and then augment the original training data with these synthetic segments. As shown in Figure \ref{fig:batch} (b), the three multi-speaker training samples are labeled as $[3/8,5/8,0]$, $[3/8,2/8,3/8]$, and $[0,4/8,4/8]$, respectively. The detailed procedure is shown in Algorithm \ref{algorithm}. Note that in Algorithm \ref{algorithm}, $T'_{nm}$ highly affects the augmentation rate (as shown in Table \ref{tab:call_der_ar}), i.e., the ratio of the number of batches that contains multiple speakers to the number of overall batches. The less $T'_{nm}$ is apt to make a larger augmentation rate.

\begin{figure}[!t]
\centering
\includegraphics[width=0.49\textwidth]{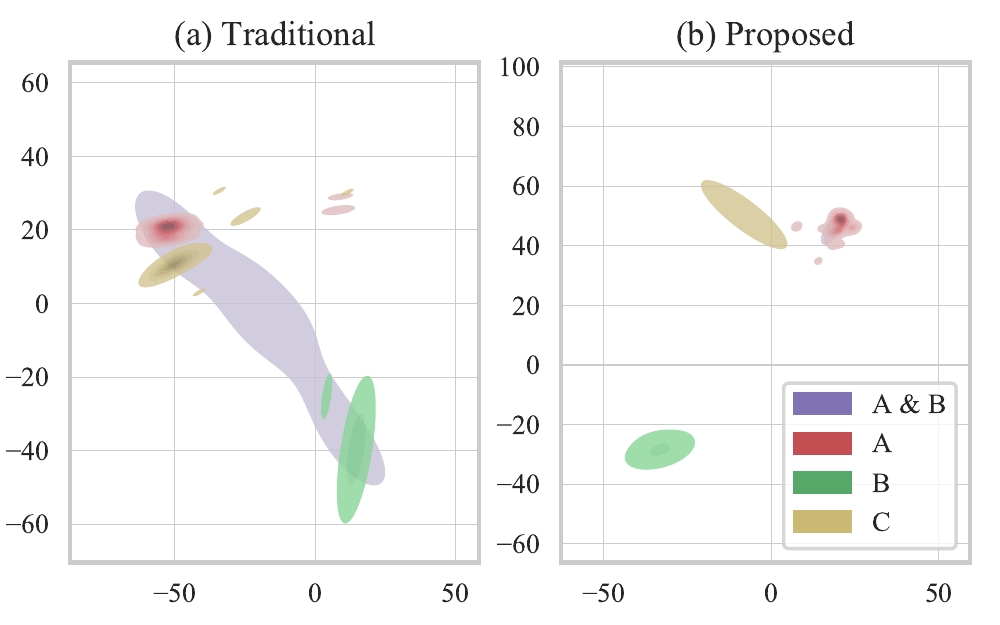}
\vspace{-20pt}
\caption{Kernel density estimate (KDE) plots of embeddings of speech segments in the CALLHOME task by t-SNE, where ``A \& B'' denotes segments that contain Speaker A and Speaker B.}
\label{fig:tsne}
\vspace{-10pt}
\end{figure}

We visualized the embeddings of four types of speech segments selected randomly from the CALLHOME task by t-Distributed Stochastic Neighbor Embedding (t-SNE) \cite{Maaten2008}. The t-SNE maps the 256-dimensional speaker embeddings into a 2-dimensional plane. From Figure \ref{fig:tsne}, we have several observations: 1) Our training scheme makes the embeddings of single-speaker segments of a specific speaker more concentrated, which means its discriminative power increases; 2) The points of ``A \& B'' are more scattered in the traditional scheme, especially not less overlapping with Speaker C, which is irrelevant with A and B. This implies that the speaker model cannot deal with segments mixed with A and B. However, in our proposed scheme, the points of ``A \& B'' are much closer to either Speaker A or Speaker B , which means our system is less likely to classify those segments as other speakers, such as Speaker C. (Note that the points of ``A \& B'' in Figure \ref{fig:tsne}(b) are too centralized so that they seem hiding behind the primary red region of ``A''. In this case, the points of ``A \& B'' are treated as those of Speaker A.) The result shows that our training scheme should be more effective than the conventional training scheme.

\begin{algorithm}[t]
\caption{Forming the proposed training batches.}
\SetAlgoLined
\KwIn{$B$: the batch size; $T$: the temporal length; $D$: the feature size; Speech of $N$ speakers}
\KwOut{The training batch $X'_{batch}$ for each training epoch}
\Begin{
$X_{batch} \longleftarrow \emptyset$\\
\While{$|X_{batch}| < T\times B$}{
Sample a speaker $S_n$ from the speaker set $\{S_i,i=1...N\}$;\\
Sample an utterance $U_{nm}$ of the speaker $S_n$;\\
Extract the $D$-dimensional frames $X_{nm}$ from $U_{nm}$;\\
Crop a suitable region $X'_{nm}$ from $X_{nm}$, where the number of $D$-dimensional frames of $X'_{nm}$ is $T'_{nm}$;\\
$X_{batch} \longleftarrow X_{batch} \bigoplus X'_{nm}$ ($\bigoplus$: concatenation along the temporal axis);\\
}
Reshape $X_{batch}$ to a tensor $X'_{batch}$ with the dimensionalities of ($B$, $T$, $D$);\\
Return $X'_{batch}$;\\
}
\label{algorithm}
\end{algorithm}

In traditional training with mini-batch gradient descent, each sample in a batch comes from a fixed set of training utterances. The training process makes the model more familiar with the batches where the speaker information in a sample is homogeneous. In contrast, our training scheme brings inter-speaker information into a sample, and such ``unseen-like'' samples are only synthesized from the original dataset without using any external resources. Augmenting the training data with these heterogeneous samples can improve the discriminative ability of the speaker model.

\section{Experiments}
\label{experiments}
\subsection{Datasets}
CALLHOME (LDC2001S97\footref{LDC2001S97}) is a widely used telephone dataset that contains conversations in multiple languages. There are 500 conversations recorded at the sampling rate of 8k. Each conversation contains 2 to 7 speakers. To train the speaker models for the CALLHOME task, the Switchboard corpus and a collection of NIST-SRE datasets, including SRE'04-06 and SRE'08, were used. The total training utterances contained 6,114 speakers for about 5,484 hours. Because CALLHOME lacks an official development set, we followed the instructions in the Kaldi recipe to equally divide the training set into two parts. CALLHOME-1 and CALLHOME-2 take turns being development and evaluation sets.

CHiME is a series of speech separation and recognition challenge. The $6^{th}$ CHiME challenge (CHiME-6) aims to deal with the problem of distant multi-microphone conversational speech diarization and recognition in the scenario of dinner party \cite{Watanabe2020}. There are two tracks: multi-array speech recognition (Track 1) and multi-array diarization and recognition (Track 2). We only discuss Track 2 in this paper because Track 1 does not involve speaker diarization. The dataset of CHiME-6 is the same as CHiME-5, and contains English conversations recorded at the sampling rate of 16k. Each conversion was recorded by 4 speakers with 4 binaural microphones wore by the speakers and 6 Microsoft Kinect microphone array devices spreading in a house. In the training set, there are 16 conversations, about 40 hours in total. In the development and evaluation sets, there are 2 conversations in each set and only the recordings of the 6 Kinects are available. For training the speaker models for the CHiME-6 task, the VoxCeleb-2 corpus \cite{SonChung2018} was used, which contains 7,351 speakers, about 2,794 hours in total.

\begin{table}
\caption{DERs on CALLHOME obtained by adding multi-speaker speech segments to the training data in different proportions. The augmentation rate means the ratio of the number of the augmented segments to the number of the total segments.}
\vspace{-5pt}
\label{tab:call_der_ar}
\centering
\begin{tabular}{cc}
\toprule
\bf{Augmentation Rate (\%)} & \bf{DER (\%)} \\ 
\midrule\midrule
42.82 & 9.33 \\
29.62 & 9.43 \\
20.84 & 9.61 \\
16.01 & 9.40 \\
12.78 & \bf{9.03} \\
\midrule\midrule
Baseline 1 (B1, X-vector) & 10.33 \\
Baseline 2 (B2, ResNet-34) & 11.26 \\
\bottomrule
\end{tabular}
\vspace{-5pt}
\end{table}

\subsection{System setup}
The two baseline systems and the proposed system were all built using the Kaldi recipe. The only difference between these systems is the speaker model. We built the first baseline system by following the Kaldi recipe, and built the second baseline system by replacing the x-vector speaker model with the ResNet-34 speaker model implemented with PyTorch. For the second baseline system, the ResNet-34 speaker model was trained in the same way as the x-vector speaker model used in the first baseline system. For the proposed system, the ResNet-34 speaker model was trained with the proposed training scheme.

In CALLHOME, the speaker models and PLDA were trained with the collection of multiple SRE datasets without any data augmentation. The input features for the x-vector model and VBD were 23-dimensional MFCCs. The parameters of VBD, such as the minimum duration, loop probability, downsampling factor, and maximum number of iterations were set to 1, 0.9, 25, and 10, respectively.

In CHiME-6, the speaker models and PLDA were trained with the VoxCeleb-2 corpus, and the ASR model was trained with the training set of CHiME-6. The VoxCeleb-2 corpus was augmented with the noise data extracted from the training set of CHiME-6 and reverberation simulated by the room impulse response (RIR) method. All the data of the development set and the evaluation set were pre-processed with the weighted prediction error (WPE) algorithm \cite{Drude2018} for de-reverberation and the BeamformIt algorithm \cite{Anguera2007} for combining all channels of all Kinects into one channel. The input features for the x-vector model and VBD were 30-dimensional MFCCs. The parameters of VBD, such as the minimum duration, loop probability, downsampling factor, and maximum number of iterations were set to 1, 0.998, 1, and 1, respectively. 

For both CALLHOME and CHiME-6, parameters related to the PLDA threshold and VBD were optimally determined by their corresponding development sets.

\subsection{Results}
\subsubsection{The CALLHOME task}

\begin{table}
\caption{Comparison of DERs on CALLHOME with and without VBD-based resegmentation.}
\vspace{-5pt}
\label{tab:call_der_reseg}
\centering
\begin{tabular}{cccc}
\toprule
\textbf{System} & \bf{DER (\%)} & \bf{+VB-diarization} & \bf{Rel. (\%)} \\ 
\midrule\midrule
Proposed & \bf{9.03} & \bf{6.87} & 23.92 \\
Baseline 1 (B1) & 10.33 & 7.89 & 23.62 \\
Baseline 2 (B2) & 11.26 & 8.57 & 23.89 \\
\midrule
Impro. over B1 (\%) & 12.58 & 12.93 & - \\
Impro. over B2 (\%) & 19.80 & 19.84 & - \\
\bottomrule
\end{tabular}
\vspace{-0pt}
\end{table}

In the CALLHOME task, we used the diarization error rate (DER) \cite{Fiscus2006} as the evaluation metric. We first evaluated the proposed training scheme by adding multi-speaker speech segments to the training data in different proportions. The results are shown in Table \ref{tab:call_der_ar}. Several observations can be drawn from the table. First, Baseline 2 (based on the ResNet-34 speaker model) is slightly worse than Baseline 1 (based on the x-vector speaker model). Second, by applying the proposed training scheme to the ResNet-34 speaker model, the resulting speaker diarization systems are always better than the baseline system based on the ResNet-34 speaker model trained in the traditional way (i.e., Baseline 2). Third, it is not that adding more multi-speaker speech segments into the training data will result in better results, but an appropriate amount. In this task, the best ratio is 12.78\%. In this case, the proposed system can reduce the DER by 12.58\% and 19.80\% compared to Baseline 1 and Baseline 2, respectively. Fourth, although Baseline 2 is worse than Baseline 1, the proposed systems built on top of Baseline 2 are always better than Baseline 1. This result confirms the effectiveness of the proposed speaker model training method. 

Next, we evaluated VBD-based resegmentation. As shown in Table \ref{tab:call_der_reseg}, all the systems can be improved by VBD-based resegmentation, with a relative DER reduction of about 24\%. The proposed system with VBD-based resegmentation outperforms the two baseline systems with VBD-based resegmentation, and can reduce the DER by 12.93\% and 19.84\%, respectively. With or without VBD-based resegmentation, the improvements over the baseline systems are about the same. The result shows that our speaker model training scheme and VBD-based resegmentation have a good combination effect.

\begin{table}
\caption{Results of the CHiME-6 development set.}
\vspace{-5pt}
\label{tab:chime_dev}
\centering
\begin{tabular}{cccc}
\toprule
\bf{System} & \bf{DER (\%)} & \bf{JER (\%)} & \bf{WER (\%)} \\ 
\midrule\midrule
Proposed & \bf{56.77} & \bf{60.62} & \bf{75.78} \\
Baseline 1 (B1) & 59.66 & 66.64 & 81.36 \\
Baseline 2 (B2) & 60.85 & 69.32 & 81.46 \\
Baseline of CHiME-6 (BC) & 63.42 & 70.83 & 84.25 \\
\midrule
Impro. over B1 (\%) & 4.84 & 9.03 & 6.86 \\
Impro. over B2 (\%) & 6.71 & 12.55 & 6.97 \\
Impro. over BC (\%) & 10.49 & 14.41 & 10.05 \\
\bottomrule
\end{tabular}
\vspace{-10pt}
\end{table}

\subsubsection{The CHiME-6 task}

In the CHiME-6 task, we compared different speaker diarization systems in terms of 3 metrics, including the DER, Jaccard error rate (JER) \cite{Ryant2019}, and word error rate (WER). DER and JER were used to evaluate the diarization result. WER was used to evaluate the downstream ASR result. We added the baseline speaker diarization system of the CHiME-6 challenge as another baseline system for comparison. We also conducted downstream ASR experiments by applying the ASR system provided by the organizer of the CHiME-6 challenge to the best diarization result of each speaker diarization system.

We first performed experiments on the development set to determine the best setting for each speaker diarization system, including the ratio of augmenting the speaker model training data with multi-speaker speech segments for the proposed system and the parameters of the VBD method. VBD-based resegmentation was applied to the results of all speaker diarization systems. The best results are shown in Table \ref{tab:chime_dev}. It is clear that the proposed speaker diarization system outperforms all the three baseline systems in terms of both DER and JER. We can also see that better speaker diarization results lead to lower ASR error rates. 

Next, we compared all the systems on the evaluation set. For all the systems, the associated parameters were determined according to the best results of the development set. The results are shown in Table \ref{tab:chime_eval}. We can see that the trend is the same as that of the results of the development set (cf. Table \ref{tab:chime_dev}).

\section{Conclusions}
\label{conclusions}
In this paper, we have proposed a novel scheme for training the speaker model to reduce the mismatch when extracting the speaker representations for performing speaker diarization. Our scheme aims to make the speaker model better handle segments containing multiple speakers, so as to get a more reliable speaker representation from each speech segment. The solution is realized by synthetically augmenting the training data for the speaker model with an appropriate number of speech segments containing multiple speakers. The results of experiments conducted on two benchmark corpora show that the speaker diarization system integrated with the proposed speaker modeling scheme is superior to the corresponding baseline speaker diarization system with conventional speaker modeling and other two baseline speaker diarization systems. The results also show that better diarization results given by the proposed systems lead to better ASR performance. In summary, the proposed speaker model training scheme is very simple but effective. Inspired by the concept of guided learning in \cite{Lin2020}, our future work will focus on how to make better use of speaker-change information, such as letting the output of the last ResNet-based layer (prior to temporal pooling) contain information to distinguish the temporal speaker-change characteristics.

\begin{table}
\caption{Results of the CHiME-6 evaluation set.}
\vspace{-5pt}
\label{tab:chime_eval}
\centering
\begin{tabular}{cccc}
\toprule
\bf{System} & \bf{DER (\%)} & \bf{JER (\%)} & \bf{WER (\%)} \\
\midrule\midrule
Proposed & \bf{59.17} & \bf{63.40} & \bf{73.54} \\
Baseline 1 (B1) & 60.97 & 65.14 & 75.10 \\
Baseline 2 (B2) & 63.51 & 69.69 & 76.59 \\
Baseline of CHiME-6 (BC) & 68.20 & 72.54 & 77.94 \\
\midrule
Impro. over B1 (\%) & 2.95 & 2.67 & 2.08 \\
Impro. over B2 (\%) & 6.83 & 9.03 & 3.98 \\
Impro. over BC (\%) & 13.24 & 12.60 & 5.65 \\
\bottomrule
\end{tabular}
\vspace{-10pt}
\end{table} 

\newpage

\bibliographystyle{IEEEtran}
\bibliography{references.bib}
\end{document}